\documentclass[twocolumn, amsmath,amssymb,aps, prl, showkeys,showpacs, superscriptaddress, floatfix] {revtex4-2}
\usepackage[colorlinks=true,breaklinks=true,allcolors=blue]{hyperref}
\usepackage{graphicx}
\usepackage{bm}
\usepackage{slashed}
\usepackage{times}

\newcommand{\ket}[1]{\ensuremath{\left| #1 \right>}}

\allowdisplaybreaks
\newcommand{\ri}{{\rm i}}
\newcommand{\re}{{\rm e}}

\begin{document}
\title{Signatures of Topological Symmetries on a Noisy Quantum Simulator}
\author{Christopher Lamb}
\email{cdl92@physics.rutgers.edu}
\affiliation{Department of Physics and Astronomy, Rutgers University, Piscataway, NJ 08854-8019 USA}
\author{Robert M. Konik}
\affiliation{Division of Condensed Matter Physics and Material Science, Brookhaven National Laboratory, Upton, NY 11973-5000, USA}
\author{Hubert Saleur}
\affiliation{Institut de physique théorique, CEA, CNRS, Université Paris-Saclay, France}
\affiliation{Physics Department, University of Southern California, Los Angeles, USA}
\author{Ananda Roy}
\email{ananda.roy@physics.rutgers.edu}
\affiliation{Department of Physics and Astronomy, Rutgers University, Piscataway, NJ 08854-8019 USA}
\begin{abstract}
        Topological symmetries, invertible and otherwise, play a fundamental role in the investigation of quantum field theories. Despite their ubiquitous importance across a multitude of disciplines ranging from string theory to condensed matter physics, controlled realizations of models exhibiting these symmetries in physical systems are rare. Quantum simulators based on engineered solid-state devices provide a novel alternative to conventional condensed matter systems for realizing these models.  
    In this work, eigenstates of impurity Hamiltonians and loop operators associated with the topological symmetries for the Ising conformal field theory in two space-time dimensions are realized on IBM's \texttt{ibm\_kingston}  simulator. The relevant states are created on the quantum device using a hybrid quantum-classical algorithm. The latter is based on a variation of the quantum approximate optimization algorithm ansatz combined with the quantum natural gradient optimization method. Signatures of the topological symmetry are captured by measuring correlation functions of different qubit operators with results obtained from the quantum device in reasonable agreement with those obtained from classical computations. The current work demonstrates the viability of noisy quantum simulators as platforms for investigating low-dimensional quantum field theories with direct access to observables that are often difficult to probe in conventional condensed matter experiments. 
\end{abstract}
\maketitle
Topological symmetries in quantum field theories are generalizations of global symmetries~\cite{Gaiotto:2014kfa} which do not necessarily obey group-like composition law and can even be non-invertible. The discovery of topological symmetries has shed new light on anomalies and renormalization group~(RG) flows in non-abelian gauge theories~\cite{Choi:2021kmx, Kaidi2022} and the standard model~\cite{Choi:2022jqy}. In contrast to their higher dimensional counterparts, topological symmetries in conformal field theories~(CFTs) residing in two space-time dimensions~\cite{Petkova:2000ip, Frohlich2006} have explicit lattice realizations in terms of anyonic~\cite{Aasen2016, Aasen:2020jwb, Belletete2020, Sinha:2023hum, Belletete2023} and quantum rotor~\cite{Roy2024} chains. This allows quantitative characterization of entire RG  flows~\cite{Kormos:2009sk, Tavares:2024vtu} and entanglement measures~\cite{Roy2024, Roy2021a}. These quantum field theories not only serve as toy models for their higher dimensional counterparts, but also, in the Hamiltonian picture, realize variations of multi-channel Kondo models~\cite{Ludwig1994,Bachas_2004} relevant for impurity problems in condensed matter physics. 

\begin{figure}
    \includegraphics[width=0.5\textwidth]{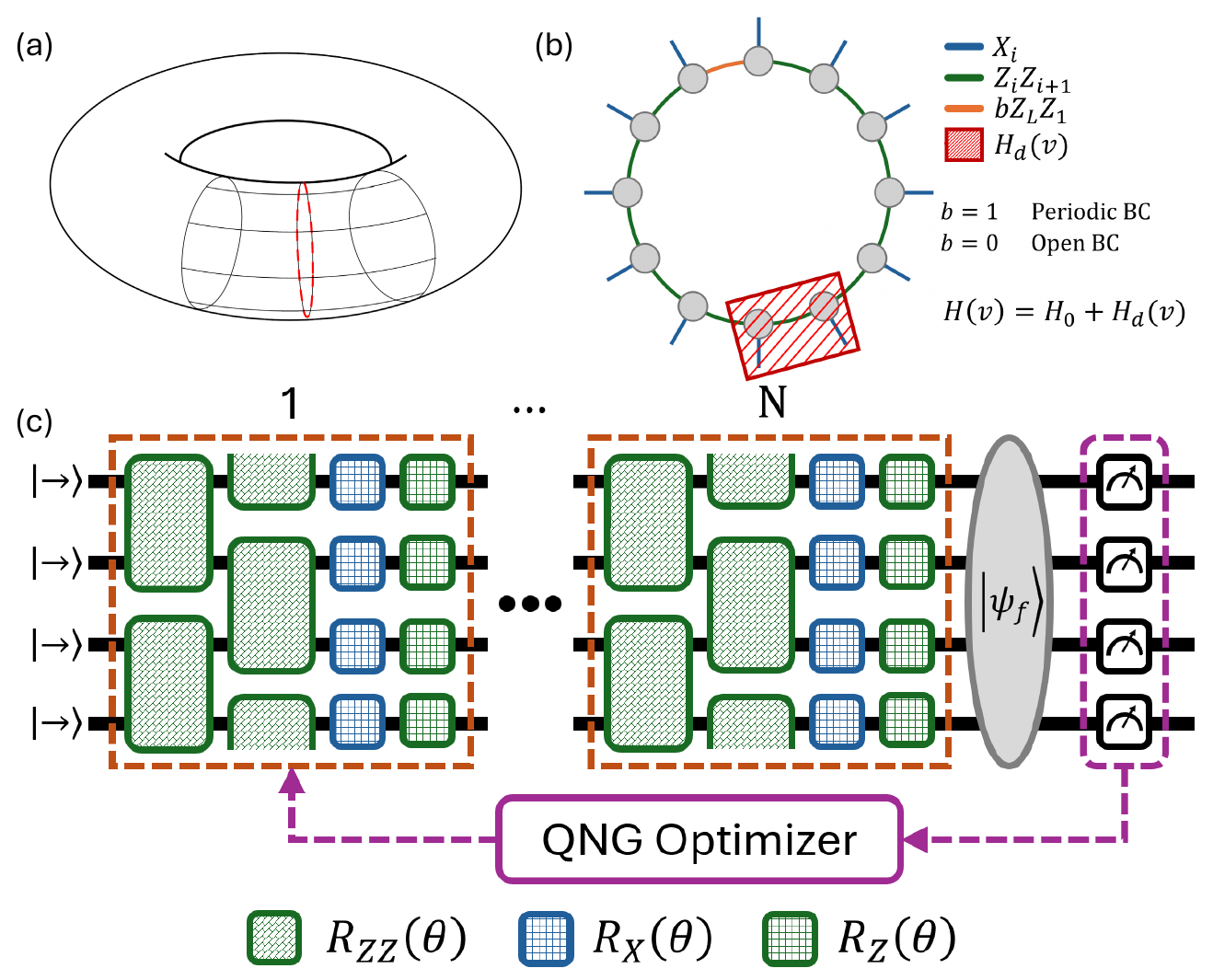}
    \caption{\label{fig_1} (a) A topological symmetry/defect line (red dashed line) in a 2D CFT on a torus. (b) Schematic of the impurity Hamiltonian for the Ising case. The green and blue lines correspond to the ferromagnetic interaction and the transverse field for the qubits~(gray circles) respectively. The orange line indicates a variable ferromagnetic coupling parameterized by~$b$ with~$b = 1(0)$ corresponding to the periodic~(open) chain. The red hatched box corresponds to the impurity part of the  Hamiltonian, $H_d(v)$, at site $j$ and $j+1$~[see Eqs.~(\ref{eq:H}, \ref{eq:H_d})]. (c) Representation of the variational quantum circuit optimization scheme used to realize the ground states of the Hamiltonian in Eq.~\eqref{eq:H}. After initialization of the qubits in the $\ket{\rightarrow}^{\otimes L}$ state, $N$ layers of the unitary operators (orange dashed boxes) are applied. The parameters of the circuit are iteratively optimized using the quantum natural gradient~(QNG) optimization method. }
\end{figure}

Despite their ubiquitous importance, controlled realizations of these symmetries in realistic physical systems are rare. Engineered quantum devices provide an alternative to the established condensed matter setups for probing these models. For 2D CFTs, lattice realizations of the relevant impurity Hamiltonians as well as the associated topological symmetry operators have already been obtained~\cite{Koo_1994, Aasen2016, Aasen:2020jwb, Belletete2018, Belletete2020, Sinha:2023hum} alongside a recipe for mapping the latter to qubit registers~\cite{Roy:2024xdi}. Even more important, certain topological symmetries in 2D CFTs are realized {\it exactly} in lattice realizations. This removes the need to realize large system-sizes to obtain agreement with field theoretic predictions -- a crucial feature allowing investigation of these symmetries using current quantum devices with modest sizes and coherence properties. In addition, the lattice models considered here are a part of an integrable family of models~\cite{Andrews:1984af}. This enables analytical computation of various equilibrium characteristics precious for comparison with experimental results. As such, these models are ideal testbeds for realization of topological symmetries on current engineered quantum systems. In fact, the latter provide direct access to several observables which are difficult to probe in conventional condensed matter experiments. 

In this work, the ground state of an impurity Hamiltonian corresponding to the non-invertible topological symmetry of the Ising CFT is realized on IBM's superconducting circuit-based \texttt{ibm\_kingston} simulator. The relevant Hamiltonian is given by~\cite{Tavares:2024vtu}
\begin{equation}
\label{eq:H}
    H(v) = -\sum_{i = 1}^{L -1}Z_i Z_{i+1} - \sum_{i = 1}^LX_i - bZ_{L}Z_{1} + H_d(v),
\end{equation}
where
\begin{equation}
\label{eq:H_d}
    H_d(v)\! = \!\frac{2\sinh^2(v)}{\cosh(2v)}(Z_jZ_{j+1} + X_j) + \frac{\sinh(2v)}{\cosh(2v)}Y_jZ_{j+1}.
\end{equation}
In the above, the operators~$X_j, Z_j$ denote the Pauli operators at the~$j^{\rm th}$ site with~$Y_j = iX_jZ_j$. The parameter~$b$ is chosen to be either 1 or 0 to switch between a periodic and an open chain~(see Fig.~\ref{fig_1}). The impurity is located between sites~$j$ and~$j + 1$ and its strength is parameterized by~$v$.  The latter tunes the strength of the impurity with~$v\rightarrow\infty$ corresponding to the Ising chain with the non-invertible~(Kramers-Wannier) duality defect ~\cite{Oshikawa1997, Grimm2001}. The case~$v = 0$ corresponds to the usual Ising chain Hamiltonian with periodic~($b = 1$) or open~($b = 0$) boundary conditions. Variation of the `Kondo screening length', given by~$l_B = e^{4v}$~\footnote{In fact, this model is closely related to the two-channel Kondo model at the Toulouse point~\cite{Emery1992}.}, relative to the length scales under investigation provides access to the entire RG flow connecting the duality and the identity defects in the Ising CFT~\cite{Tavares:2024vtu}. 

The characteristics of the aforementioned model along the RG trajectory can be obtained by computing the scaling of the ground state energy as a function of the dimensionless parameter~$L/l_B$. Using standard arguments~\cite{Cardy:1986ie, Petkova:2000ip}, the relevant scaling dimension for~$v \rightarrow \infty(0)$ is given by~$1/16(0)$ with numerical results available for the intermediate values~\cite{Tavares:2024vtu}. In addition, a particularly interesting feature of the duality defect Hamiltonian is that it couples the order fields to the disorder fields on the two sides of the defect~\cite{Oshikawa1997}. This leads to a dramatic change in the behavior of the correlation function~$\langle Z_1Z_{r}\rangle$ as the site index $r$ crosses the defect location. Here, the correlation function is computed with respect to the ground state of the Hamiltonian for a given value of~$v$. For~$v = 0$, the correlation function exhibits the usual power-law decay with the well-known exponent governed by the scaling dimension of the spin field of the Ising CFT. However, for~$v\rightarrow\infty$, the correlation drops abruptly to zero as soon as~$r$ crosses the defect location. These properties, which are zero temperature equilibrium characteristics of the model, are difficult to measure in a typical condensed matter experiment. However, the relevant ground states can be created in a quantum simulator by the application of a suitable parametrized quantum circuit with both periodic and open boundary conditions straightforwardly realizable.  Subsequently, the ground state energy and the necessary two-point correlation function can be obtained by performing measurements of one and two-qubit observables. This is described next. 

\begin{figure}
    \includegraphics[width=0.5\textwidth]{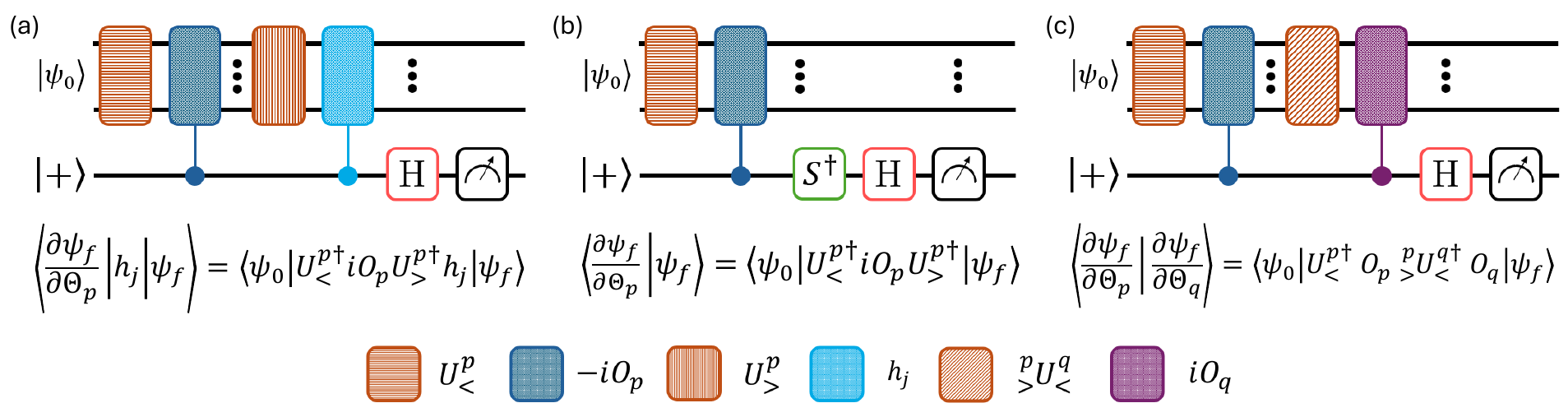}
    \caption{\label{fig_2n} Circuits for the evaluation of the gradient of the cost function~[panel (a)] and the Fubini-Study matrix element~[panels (b), (c)] required for the QNG optimization update~[Eq.~\eqref{eq:QNG_update}]. (a) For the computation of the gradient~$\partial{\cal L}/\partial\Theta_p$, first the portion of the circuit until the rotation by~$\Theta_p$~(denoted by $U^p_<$) is applied to the qubits initialized to~$|\psi_0\rangle$. This is followed by a controlled-$\tilde{O}_p$ rotation with an ancilla qubit, initialized to~$|+\rangle$, as control and the remaining gates of the circuits~(denoted by~$U^p_>$). Here,~$\tilde{O}_p = -iO_p$. Finally, a controlled unitary rotation is performed by the~$j^{\rm th}$ term of the Hamiltonian,~$h_j$. Averaging over the X-measurements of the ancilla qubit yields the contribution to the gradient from the~$j^{\rm th}$ term. The total gradient is the sum of the different such contributions. (b) Computation of the overlaps~$\langle\partial\psi_f/\partial\Theta_p|\psi_f\rangle$. In this case, after the application of~$U^p_<$ and the controlled-$\tilde{O}_p$, average is performed over the Y-basis measurements of the ancilla qubit. (c) Computation of the overlaps~$\langle\partial\psi_f/\partial\Theta_p|\partial\psi_f/\partial\Theta_q\rangle$. After application of~$U^p_<$ and controlled-$\tilde{O}_p$, the portion of the circuit until the rotation by the angle~$\Theta_q$ is applied~(denoted by~${}^p_>U^q_<$), followed by a controlled-$iO_q$. Averaging over the X-measurement results yields the relevant overlap. See Secs. S1 and S2 of the Supplementary Material for more details.}
\end{figure}

To realize the relevant ground state, a parameterized circuit ansatz is used which is a variation of the quantum approximate optimization algorithm operator ansatz~\cite{Farhi2014, Lloyd2018, Hadfield2019, Morales2020, Roy2023efficient}. Starting with all qubits pointing along the $|\rightarrow\rangle$ state, $N$ layers of unitary operators (orange boxes in Fig.~\ref{fig_1}) are applied. The final state~$|\psi_f\rangle$ is given by:
\begin{align}
\label{eq:QC_1}
|\psi_f\rangle &= U^NU^{N-1}\cdots U^1, \ U^\alpha = U_{Z}^{\alpha}U_{X}^{\alpha}U^\alpha_{ZZ}, \nonumber\\
U_X^{\alpha} &= \prod_{j}R_{X_j}(\zeta^\alpha_j),\ U_Z^{\alpha} = \prod_{j }R_{Z_j}(\phi^\alpha_j),\nonumber\\ U^\alpha_{ZZ} &= \prod_{j}R_{Z_jZ_{j+1}}(\theta^\alpha_j),
\end{align}
where~$\alpha(j)$ is the layer~(site) index and~$R_{\rm O}(\varphi)$ is the unitary rotation by angle~$\varphi$ with generator~$O$. Here, the boundary condition of the circuit ansatz is taken to be identical to the boundary condition imposed on the target Hamiltonian~[Eq.~\eqref{eq:H}], although this is not essential for the approach to work~(see Refs.~\cite{Rogerson2024, Roy2024ec} for more general ans\"{a}tze). Note that the circuit ansatz does not preserve the~$\mathbb{Z}_2$ symmetry of the Hamiltonian arising from the conserved operator~$\prod_j X_j$. The parameters,~$\theta_j^\alpha, \phi_j^\alpha$ and~$\zeta_j^\alpha$ are subsequently determined iteratively using the quantum natural gradient~(QNG) method~\cite{Stokes2020, Wierichs2020, Roy2023efficient}.  The latter is a sophisticated variation of the conventional gradient descent methods like BFGS~\cite{Fletcher2013} or ADAM~\cite{Kingma2017}, where the optimization is performed by taking into the account the geometry of the manifold of quantum states~\cite{Zanardi2007, Kolodrubetz2013, Kolodrubetz2017}. The central ingredient is the Fubini-Study metric tensor whose elements are given by~$\mathbf{g}_{pq}=\Re(G_{pq})$, where
\begin{align}\label{eq:G_def}
G_{pq}(\vec{\Theta})\! &= \!\Big\langle \frac{\partial\psi_f}{\partial \Theta_p}\Big| \frac{\partial\psi_f}{\partial \Theta_q}\Big\rangle - \Big\langle \frac{\partial\psi_f}{\partial \Theta_p}\Big| \psi_f\Big\rangle\Big\langle \psi_f\Big| \frac{\partial\psi_f}{\partial \Theta_q}\Big\rangle
\end{align}
and~$\vec{\Theta}$ is a vector of all the circuit parameters to be determined. From Eq.~\eqref{eq:QC_1}, the number of such parameters is~$3LN$ and $(3L - 1)N$ for~$b = 1$ and~$b = 0$ respectively. Then, the circuit parameters at the~$(t+1)^{\rm th}$ step are given by:
\begin{equation}
\label{eq:QNG_update}
\vec{\Theta}_{t+1} = \vec{\Theta}_{t} - \eta\mathbf{g}^{-1}\frac{\partial{\cal L}}{\partial\vec{\Theta}_t},
\end{equation}
where~${\cal L} = \langle\psi_f|H|\psi_f\rangle$ is the relevant cost-function being minimized to obtain the ground state and~$\eta$ is the learning rate. The QNG-based optimization approach often outperforms its competitors~\cite{Wierichs2020, Roy2023efficient, Roy2024ec}. This superior efficacy comes at the price of additional quantum circuits that need to be run on the quantum simulator. This is because a computation of~$\mathbf{g}_{pq}$ is required at each optimization step in addition to the gradient of the cost function~[Eq.~\eqref{eq:QNG_update}]. In contrast to the existing proposals for the determination of the Fubini-Study metric using a parameter-shift rule~\cite{Stokes2020, Killoran2021, Wierichs:2021nwf}, here an alternative is proposed for the computation of the gradients and the metric elements needed for the optimization process.  In this scheme~(Fig.~\ref{fig_2n}), the relevant multi-point correlation functions are computed by applying suitable controlled-unitary operations followed by measurements of the ancilla qubit in the X and Y bases. As shown in Fig.~\ref{fig_2n}, the circuits required to compute~$\mathbf{g}_{pq}$ are similar in depth as the gradient evaluations, but  lead to an increase in the total number of circuits evaluated on the quantum processor. The measurement-based method using the ancilla qubit is verified to asymptotically agree with exact computations as the number of measurement shots is increased~(see Supplementary Materials, Secs. S1 and S2, for details).

\begin{figure}
    \centering
    \includegraphics[width=0.5\textwidth]{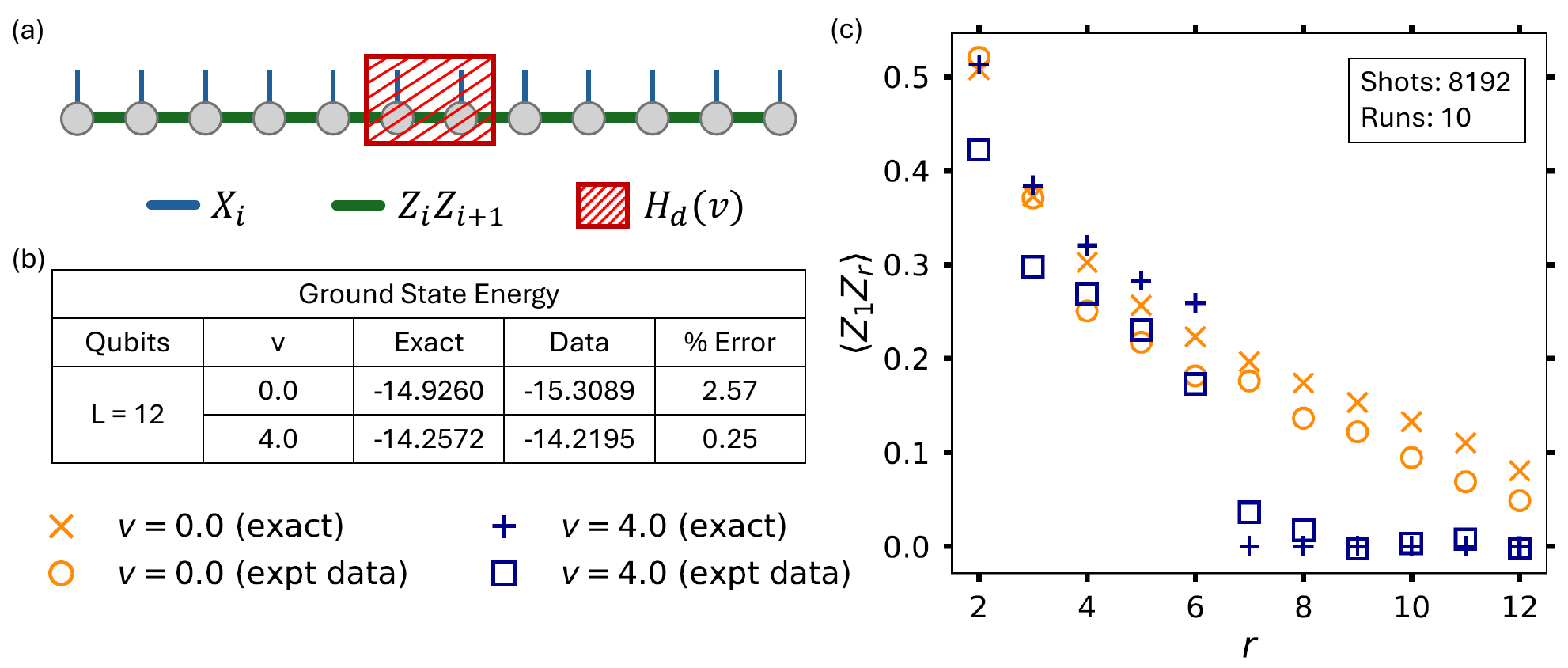}
    \caption{\label{fig_2} (a) Schematic of an open Ising chain with 12 qubits and the impurity between sites~$j, j+1$ with~$j = 6$. The topological defect is introduced (removed) by tuning the parameter~$v$~[Eq.~\eqref{eq:H}] to $\infty(0)$. (b) Ground state energies obtained from measurement of single and two-qubit correlation functions corresponding to the different terms of the Hamiltonian~[Eq.~\eqref{eq:H}] using ZNE and 5 runs, each with 1024 shots. For comparison, exact results are also shown. (c) Raw measurement data for the correlation function~$\langle Z_1Z_r\rangle$ for a 12-qubit chain with~$v = 0$  and 4 shown using orange circles and blue squares respectively. For comparison, the results computed using exact diagonalization are also shown. In contrast to the~$v = 0$ case where the correlation function exhibits a power-law decay characteristic of a critical theory, for~$v = 4$, the correlation function drops abruptly to zero as the defect location is traversed. The data was obtained averaging over 10 runs with 8192 shots per run. For panels (b, c), the circuit parameters for the realization of the ground state were computed classically using QNG optimization method. The so-obtained circuit was then implemented on \texttt{ibm\_kingston} followed by relevant measurements~(see main text for more details).}
\end{figure}

Next, results are presented for the Ising chain with an impurity at the center of the chain~[Fig.~\ref{fig_2}(a)]. These results were obtained from the 156-qubit \texttt{ibm\_kingston} simulator. Due to limited access to the quantum hardware, the circuit parameters required to realize the ground states were determined using classical computers. Subsequently, these circuits were implemented on the quantum hardware to compute the relevant observables. The number of layers required to reach an error in target energy of $<0.1\%$ was $N = L/2$. This is compatible with the observations of Ref.~\cite{Roy2023efficient} for critical spin chain Hamiltonians like that in Eq.~\eqref{eq:H}. The learning rate was chosen to be~$\eta = 0.05$. Each of the circuits was compiled into an ISA circuit with a pass manager configured with the highest level of circuit optimization and SABRE routing. After the initial compilation, each circuit was recompiled 20 times to find the circuit with the least number of two-qubit gates. The Dynamical Decoupling sequence \texttt{XpXm} and Twirled Readout Error Extinction were applied to all the results. Zero Noise Extrapolation (ZNE) was applied to all of the results with the exception of the correlation functions for the open boundary case. The ZNE strategy used a range of noise factors from 1 to 3 in steps of 0.2, the \texttt{gate\_folding\_back} amplifier and second degree polynomial extrapolator. Fig.~\ref{fig_2}(b) shows the ground state energies for an open Ising chain with 12 qubits for~$v = 0$~(no defect) and~$v = 4$~(sufficient to realize the duality defect for the chosen system size) obtained by computing the relevant one and two point correlation functions. With the help of the different error mitigation strategies, the energies are obtained to within a few percent of the exact results. Note that even though the noiseless simulations had an error~$< 0.1\%$, the noise in the actual quantum device resulted in higher errors in the data. Fig.~\ref{fig_2}(c) shows the results for the computation of the correlation function~$\langle Z_1Z_r\rangle$. In contrast to the~$v = 0$ case where~$\langle Z_1Z_r\rangle$ exhibits a power law decay characteristic of critical theories, the correlation function {\it drops abruptly to zero} as~$r$ is varied across the defect location. This can be viewed as a `smoking-gun signature' of the topological symmetry realized by the impurity Hamiltonian. Indeed, the duality defect couples the order and disorder fields of the Ising chain on either side of the defect which leads to the observed behavior~\cite{Oshikawa1997}.

\begin{figure}
    \includegraphics[width=0.5\textwidth]{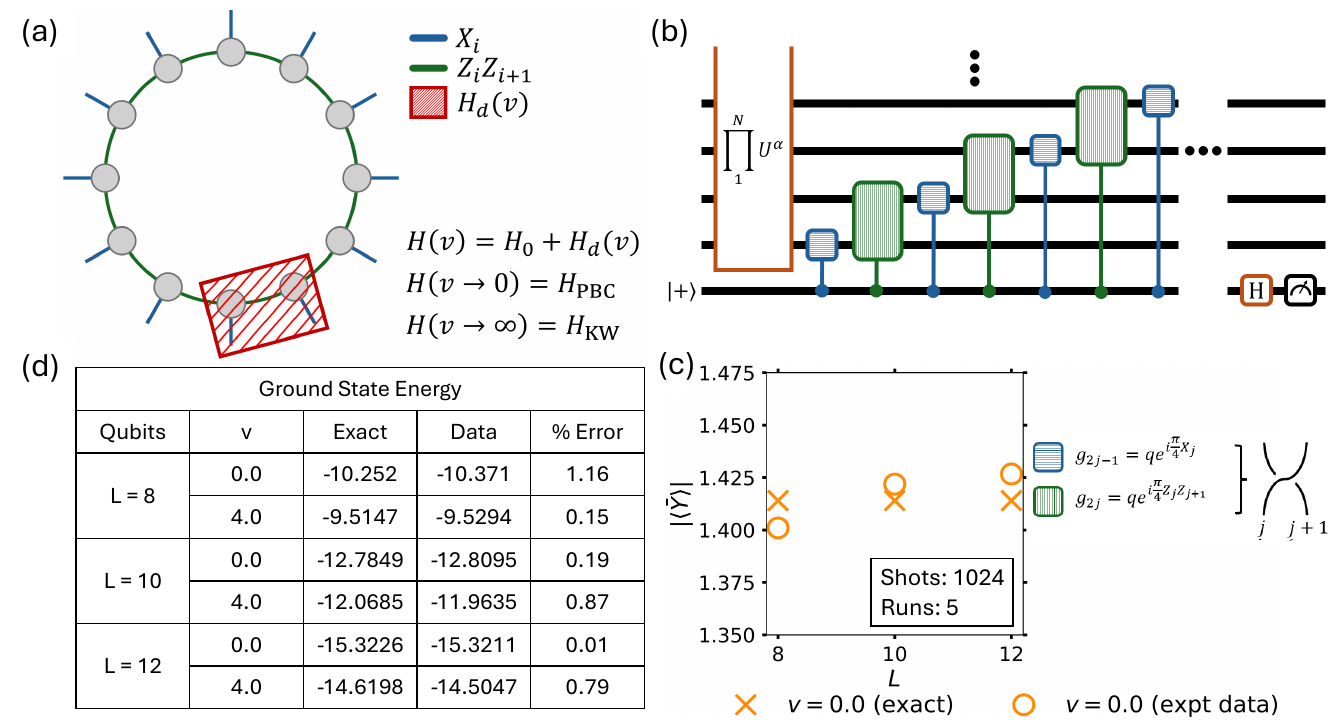}
    \caption{\label{fig_3} (a) Schematic of the Ising chain with periodic boundary conditions with an impurity between sites~$j, j+1$~[Eq.~\eqref{eq:H}]. The model reduces to the periodic (duality-twisted) Ising chain as~$v\rightarrow0(\infty)$. (b) Quantum circuit for the measurement of the topological symmetry operator, $\bar{Y}$. After preparing the qubits in the ground state of~$H(v = 0)$ using circuit parameters obtained from  QNG-based optimization, controlled braid-operators~$g_j$-s~[Eq.~\eqref{eq:g}] are applied with an ancilla qubit as the control. Averaging over X-measurement of the ancilla yields the desired expectation value. (c) Comparison of measurement results for~$|\langle\bar{Y}\rangle|$~[Eq.~\eqref{eq:Yb}] with exact predictions. The orange  circles (crosses) denote experimental data (exact results) for the expectation value of~$\bar{Y}$. (d) Comparison of the ground state energies obtained from the measurement of one and two-qubit correlation functions with exact results for~$L = 8, 10, 12$. The experimental data is obtained using ZNE and averaging over 5 runs of the experiment on the \texttt{ibm\_kingston} simulator with each run containing 1024 shots. See main text for details regarding the measurement protocol. }
\end{figure}

Topological symmetries in CFTs also leave their imprints in their respective g-functions~\cite{Affleck1991}. The difference between the g-functions can be obtained from the change in the thermodynamic entropy with temperature. The thermodynamic entropy is amenable to numerical and sometimes analytical computations. It is also the quantity often accessible in experimental settings. However, interchanging the role of space and time in the torus~[Fig.~\ref{fig_1}(a)], the different g-functions can directly be obtained from the expectation value of the loop operator in the ground state of the periodic Hamiltonian~[Fig.~\ref{fig_3}(a)]. Indeed, the ground state is an eigenstate of the topological symmetry operator with the relevant eigenvalue being the g-function. Owing to its origin in integrability, the forms of the different topological symmetry operators are known exactly in terms of the lattice spin operators~\cite{Aasen2016, Belletete2020, Sinha:2023hum}. The topological symmetry operator with a nontrivial g-function for the Ising CFT is the one corresponding to the duality defect~[$v\rightarrow\infty$ in Eq.~\eqref{eq:H}]. To evaluate the g-function using the spin chain model considered here, it is sufficient to consider the operator~\footnote{This corresponds to the dropping the operator performing translation by half a lattice site from the loop operator defined in Ref.~\cite{Belletete2020, Sinha:2023hum}. The dropped operator does not change the analyzed g-function.}
\begin{equation}
\label{eq:Yb}
\bar{Y} = (-q)^{L}g_1^{-1}\ldots g_{2L - 1}^{-1} + {\rm h.c.},
\end{equation}
with~$q = \ri \re^{\ri\pi/4}$ and the braid operators~$g_j$-s given by
\begin{equation}
\label{eq:g}
g_{2j-1} = q\re^{\ri\pi X_j/4}, g_{2j} = q\re^{\ri\pi Z_jZ_{j+1}/4}.
\end{equation}
The expectation value of~$\bar{Y}$ is obtained by first creating the ground state of the periodic Hamiltonian using the variational approach described earlier and then applying controlled unitary operators~[Fig.~\ref{fig_3}(b)] where an ancilla qubit plays the role of the control qubit. Averaging over measurements of the ancilla qubit in the X-basis yields the desired expectation value. Fig.~\ref{fig_3}(c) shows the results for~$|\langle \bar{Y}\rangle|$ (orange circles) while comparing with exact result of~$\sqrt{2}$ (orange crosses) for~$L = 8, 10, 12$. The measurement protocol is identical to that used to obtain the results shown in Fig.~\ref{fig_2}(b). Note that the relevant topological symmetry operator is realized exactly on the lattice which leads to good agreement with field theory predictions even for such small system-sizes. For benchmarking purposes, ground state energies are also obtained from the measurement of correlation functions corresponding to the different terms of the Hamiltonian~[Eq.~\eqref{eq:H}] for different systems-sizes. The measurement results, alongside those obtained from exact computations, are shown in Fig.~\ref{fig_3}(d). 

In summary, this works realizes the eigenstates of Hamiltonians and associated loop operators for topological symmetries in the Ising CFT and performs measurements of relevant observables on IBM's \texttt{ibm\_kingston} simulator. The relevant eigenstates are created on the noisy quantum simulator using a hybrid quantum-classical algorithm based on a variational quantum circuit. The parameters of the latter are determined using the quantum natural gradient optimization method. Measurements are performed for observables that capture the signatures of the non-invertible topological symmetry of the Ising CFT. The measurement results are in close agreement with those obtained using classical methods demonstrating the noise-resilience of the proposed protocol. 

In contrast to transport characteristics probed in typical condensed matter experiments, as shown in this work, hybrid algorithms on current quantum simulators provide access to a wider variety of observables for low-dimensional quantum field theories. The current work can be straightforwardly generalized using the framework developed in Ref.~\cite{Roy:2024xdi} to probe all non-invertible symmetries in minimal models of CFTs in two space-time dimensions~\cite{Sinha:2025jhh} and analyze characteristics along the RG flows connecting the various fixed points~\cite{Kormos:2009sk, Tavares:2024vtu}. With further advancement of quantum technologies, more exotic quantum field theories including those realized by non-compact~\cite{Bytsko2006} and non-hermitian spin chains~\cite{Ikhlef2012, Bazhanov:2020dlm} could be realized using quantum simulators opening the door to investigation of QFTs which lack controlled realization in other experimental setups. 

The authors thank David Rogerson and Madhav Sinha for discussions and related collaborations. This research used resources of the Oak Ridge Leadership Computing Facility, which is a DOE Office of Science User Facility supported under Contract DE-AC05-00OR22725. AR and RMK were supported by the U.S. Department of Energy, Office of Basic Energy Sciences, under Contract No. DE-SC0012704. The work of H.S. was supported
by the French Agence Nationale de la Recherche (ANR) under grant ANR-21- CE40-0003 (project CONFICA).

\bibliography{/Users/ananda/Dropbox/Bibliography/library_1}

\end{document}